\documentclass{article}
\usepackage{smc2025}
\usepackage[caption=false, font=footnotesize]{subfig}
\usepackage{paralist}
\usepackage[figure,table]{hypcap}

\usepackage[whole]{bxcjkjatype}

\usepackage{alphabeta}
\usepackage{arabtex}
\usepackage[LFE,LAE,LGR,T2A,T1]{fontenc}
\usepackage[greek, russian, main=english]{babel}

\usepackage{amsmath,amssymb,amsfonts}
\usepackage{algorithmic}
\usepackage{caption}
\usepackage{graphicx}
\usepackage{textcomp}
\usepackage{xcolor}
\usepackage{adjustbox}
\usepackage{algorithm}
\usepackage{float}

\def\papertitle{How to Infer Repeat Structures in MIDI Performances}

\author[1]{\mbox{\firstname{Silvan}\middlename{David}\lastname{Peter}}}
\author[1]{\mbox{\firstname{Patricia}\lastname{Hu}\originalname{胡紫漪}}}
\author[1, 2]{\mbox{\firstname{Gerhard}\lastname{Widmer}}}

\affil[1]{\institution{Institute of Computational Perception, Johannes Kepler University}\city{Linz}\country{Austria}\affiliationtype{University}}
\affil[2]{\institution{LIT AI Lab, Linz Institute of Technology}\city{Linz}\country{Austria}\affiliationtype{Music}}

\completesetup

\title{\papertitle}
\begin{document}

	\capstartfalse
	\maketitle
	\capstarttrue

\section{Introduction}\label{sec:introduction}
MIDI performances are generally expedient in performance research and music information retrieval, and even more so if they can be connected to a score~\cite{Peter-2023,cancino2018computational}. 
This connection is usually established by means of alignment, linking either notes or time points between the score and the performance. The first obstacle when trying to establish such an alignment is that a performance realizes one (out of many) structural versions of the score that can plausibly result from instructions such as repeats, variations, and navigation markers like ‘dal segno/da capo al coda’. 
A score needs to be unfolded, that is, its repeats and navigation markers need to be explicitly written out to create a single timeline without jumps matching the performance, before alignment algorithms can be applied. 
In the curation of large performance corpora this process is carried out manually, as no tools are available to infer the repeat structure of the performance~\cite{BukeyFD24}.

To ease this process, we develop a method to automatically infer the repeat structure of a MIDI performance, given a symbolically encoded score including repeat and navigation markers. The intuition guiding our design is: 1) local alignment of every contiguous section of the score with a section of a performance containing the same material should receive high alignment gain, whereas local alignment with any other performance section should accrue a low or zero gain. And 2) stitching local alignments together according to a valid structural version of the score should result in an approximate full alignment and correspondingly high global accumulated gain if the structural version corresponds to the performance, and low gain for all other, ill-fitting structural versions.

In Section 1 we provide a detailed description of our proposed method, and in Section 2, we demonstrate its application on the (n)ASAP dataset~\cite{Peter-2023}. The method proves a reliable and useful preprocessing tool for large-scale performance data curation and even uncovered several previously unknown errors in the dataset.

\begin{figure*}[t]
    \centering
    {\includegraphics[width=2\columnwidth]{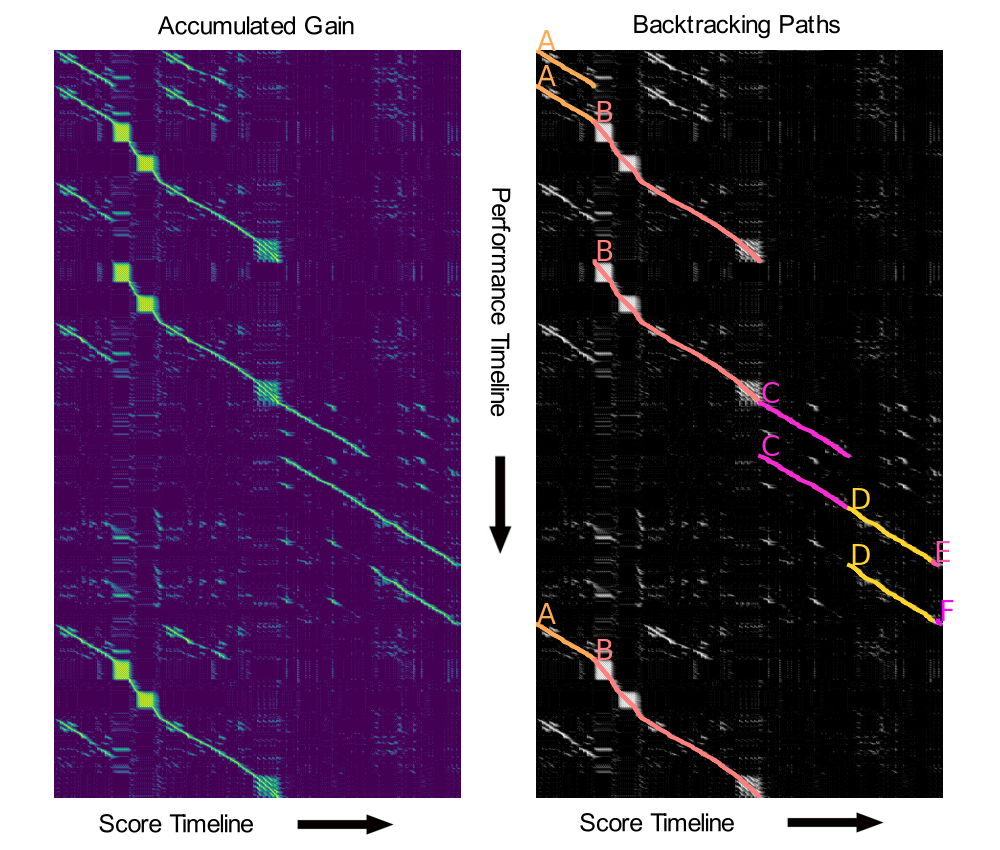}} 

    \caption{An example inference for a performance of Beethoven’s Piano Sonata No. 11, Op. 22, third movement. The left image shows the accumulated gain of the local sequence alignment, the performance is in the vertical dimension and starts at the top, the score is in the horizontal dimension. The right image superimposes the backtracked segments. This piece contains four repeated segments (A,B,C, and D), the last one ending in Volta brackets (E and F). After the “Da Capo al Fine” at the end of segment F, the performance plays each segment (A and B) only once.}
    \label{fig:match}
\end{figure*}
	
\section{Method}

The proposed method is based on three steps. First, we perform local sequence alignment to identify well-aligned segments in the performance that correspond to segments in the score. Next, we combine these local alignments into a global alignment for all possible structural versions of the score, recording their respective global alignment gain. Finally, we select the structural version that maximizes the global alignment gain.

\subsection{Local Sequence Alignment }
Local sequence alignment algorithms such as the Smith-Waterman algorithm~\cite{smith1981identification} are well-suited for identifying segment-wise correspondences. We adapt an existing symbol-level DTW-type alignment algorithm~\cite{peter23online} into a Smith-Waterman-style local alignment algorithm by incorporating match gains, mismatch penalties, and constraining the gain to non-negative values.  We use a purely pitch-based symbol-level feature representation paired with a pairwise metric for alignment which is designed to work with arbitrary polyphonic scores and performances. The score is represented as an array where each element is a set of MIDI pitches corresponding to a unique score onset: 
\begin{equation*}
Sco = [\{P_1 , \ldots ,P_n\} , \{P_1 , \ldots ,P_m\} , \{P_1 , \ldots ,P_o\}, \ldots ]
\end{equation*}
where all pitches $P \in \{0 , \ldots , 127\}$.
The performance is represented as an array of consecutive MIDI pitches:  
\begin{equation*}
Per = [ P_1 , ... ,P_n]
\end{equation*}
The local metric between a score onset pitch set and a performed pitch is then defined as:
\begin{align*}
m(i,j) = m(Per[i],Sco[j]) = m(P_i, \{P_1 , \ldots ,P_n\}) \\
 :=\begin{cases}
      1 & \text{if } P_i \in \{P_1 , \ldots ,P_n\}\\
      -1 & \text{otherwise}
    \end{cases}   
\end{align*}
Note that the features and the local metric only process pitch information. Uncommon in music alignment, the metric represents a gain and not a cost value, that is, matching pitches result in a higher value. To compute the accumulated gain from this local gain, we use a modified Smith-Waterman dynamic programming algorithm. Figure 1 shows the output of this gain accumulation. Following this visual intuition we assume the performance to be in the vertical direction (each row index corresponds to a performed MIDI pitch) and the score in the horizontal (each column index corresponds to a set of pitches at a unique score onset). The algorithm loops over all performance and score indices and the local accumulated gain is computed from one of two possible previous gains: a diagonal step, i.e., a new score pitch set and a new performance note, or a vertical step, i.e., a new performance note at the same score pitch set, can be added to the gain value. The accumulated gain $ag(i,j)$ is then computed by maximizing across the two possible steps:
\begin{equation*}
ag(i,j) = max( ag(i-1,j), ag(i-1, j-1) )+ m(i,j) 
\end{equation*}

The accumulated gain matrix $ag$ is initialized at zero. To ensure non-negative alignment scores, a standard Smith-Waterman algorithm contains a reset step:  $ac(i,j)=max(0,ac(i,j))$. We modify the accumulation bounds by setting upper and lower limits.
The maximal gain is bounded by the gain value 10, with exponentially decreasing gain increments approaching this upper bound asymptotically. The minimum gain is bounded by zero, decreasing linearly until reaching a value of 1, after which it asymptotically approaches zero.. Bounded gains and only two update directions is a peculiar setup for music alignment, yet in this case it enables distinct, clearly separated local alignment ridges in the gain accumulation matrix (see Fig. 1) which in turn simplifies the following backtracking.

\begin{table*}
 \centering
  \adjustbox{max width=\linewidth}{
\bgroup
\def\arraystretch{1.2}%
\begin{tabular}{llll}
Predicted Structure & Dataset Structure & Piece \& Performance & Explanation \\ \hline
ABC & ABBC & Beethoven Piano Sonata 17-3 KaszoS11M & skip in C, no repeat of B \\
ABDE & ABCBDE & Beethoven Piano Sonata 18-2 Schmitt01 & no repeat of B\\
ABDE & ABCBDE & Beethoven Piano Sonata 21-1 Sekino05M & no repeat of B\\
ABDE & ABCBDE & Beethoven Piano Sonata 21-1 YOO05 & no repeat of B\\
ABDE & ABCBDE & Beethoven Piano Sonata 26-3 HONG05M & no repeat of B\\
ABDE & ABCBDE & Beethoven Piano Sonata 26-3 Huang02 & no repeat of B\\
ABDE & ABCBDE & Beethoven Piano Sonata 26-3 Tysman01 & no repeat of B\\
AABCDCEFGIJ & AABCDCEFGHGIJ & Beethoven Piano Sonata 31-2 Na06 & no repeat of G\\
ABDE & ABCBDE & Beethoven Piano Sonata 32-1 Faliks01 & no repeat of B\\
ABDE & ABCBDE & Beethoven Piano Sonata 32-1 FALIKS012004 & no repeat of B\\
ABDE & ABCBDE & Beethoven Piano Sonata 32-1 Park01 & no repeat of B\\
ABDE & ABCBDE & Beethoven Piano Sonata 32-1 POTAMO01 & no repeat of B \\
ABC & ABBC & Chopin Piano Sonata 2-3 KaszoS15 & no repeat of B \\
ABC & ABBC & Liszt Hungarian Rhapsody 6 Ye04 & no repeat of B\\
ABCEFGHI & AABCDCEFGGHHI & Schumann Kreisleriana 2 JohannsonP03 & skip in F, no repeat of H\\
ABCBDEGH & ABCBDEFEGH & Schumann Kreisleriana 3 JohannsonP04 & no repeat of E\\
ABCBDEGH & ABCBDEFEGH & Schumann Kreisleriana 3 Yarden09 & no repeat of E\\
\end{tabular}
\egroup}
 \caption{Annotation errors identified in the (n)ASAP Dataset. 
 Repeat structures are given as strings where we alphabetically mark all score sections delimited by structurally relevant symbols (repeats, volta brackets, da capo, al segno, etc.). 
 The first column shows the predicted structures, manual inspections show that they are correct for all but the Kreisleriana 2 and the Beethoven Piano Sonata 17-3 performances where large unnotated skips (10+ measures) hinder both automatic matching and manual annotation.
 The second column shows the repeat structures given by the dataset, the third the composer, piece (-n identifies the nth movement), and performer. 
 The last column explains the source of the mismatch.
 }
 \label{tab:errors}
\end{table*}

\subsection{Score-informed Backtracking}
Local sequence alignment produces a gain matrix that is used to extract local matching subsequences. However, these local alignments are extracted in an unstructured fashion: backtracking from the highest gain ignores position, does not guarantee full coverage of the performance, and may not preserve sequential constraints in the score. Figure 1 shows the type of path that we would like to backtrack from the accumulated gain matrix (left) in the same matrix with superimposed correct path segments (right). Several candidate alignments are visible which violate musical assumptions, e.g., at the very top (the beginning of the performance) we see two diagonal lines which start in the middle of score segment B. This is not a valid place to begin a performance, yet standard backtracking will produce these local alignments. To address this issue, we use score-informed backtracking. First, we compute all possible, musically valid, structural versions based on repeat, variation and navigation markers (e.g., repeating at most twice, playing the coda once), and label the segments between markers alphabetically. 

For each structural version, backtracking starts at the cost matrix position corresponding to the last segment, and continues until it reaches either the start of that segment or a cost of zero, at which point  the local alignment is added to the alignment stack of that version. The alignment cost is added to the global alignment cost. The process is then repeated for the next segment, starting from the end of that segment in the score and the performance position where the previous local alignment ended. 

This continues until the first segment of the score version is aligned. The global cost is calculated as the sum of the local alignment costs, weighted by a penalty for using more segments. Finally, we select the score version that minimizes this global cost. If the number of possible score versions becomes too large due to combinatorial explosion, this step can be carried out with a smaller subset of commonly performed versions.

\section{Demonstration}

We test our method on all performances of the (n)ASAP dataset and compare against the true repeat structure of the performances. The dataset contains 110 performances with notated repeats or skips (whether actually performed or not), and our method initially returned roughly 85\% correct repeat structures for these. Upon closer inspection of the errors, we found that all 17 errors were due to faulty dataset annotations. Among those we found 15 performances where the dataset error is only due to a mismatch in repeats and thus easily fixable. However, there are also two performances which skip musical material outside of marked sections. 
See Table~\ref{tab:errors} for a detailed list of the found errors.
We did not double-check the correctness of all 93 performances where the predicted structure corresponds to the dataset annotations. 
As far as these remaining dataset annotations are to be trusted, our method is correct for all performances where a musically valid repeat structure was played and uncovered a high number of incorrect annotations.

\section{Conclusion}
In this article, we present a simple, fast, and automatic tool to infer the repeat structure in recorded MIDI performances. 
Our method builds on previous alignment models and can be used as a pre-processing or checking tool in the curation of large-scale piano performance corpora. 
It was tested and developed with solo piano performances of the (n)ASAP Dataset and in principle works with arbitrary polyphonic music by processing purely pitch-based information. 

Our method is available in the python music alignment library parangonar: 
\url{https://github.com/sildater/parangonar}

The dataset is available at: \url{https://github.com/CPJKU/asap-dataset}

\begin{acknowledgments}
This research acknowledges support by the European Research Council (ERC), under the European Union's Horizon 2020 research and innovation programme, grant agreement No.~101019375 \textit{Whither Music?}.
\end{acknowledgments}

\bibliography{refs}
	
\end{document}